\documentstyle[epsf,aps]{revtex}

\def\be{\begin{equation}}
\def\ee{\end{equation}}
\def\bea{\begin{eqnarray}}
\def\eea{\end{eqnarray}}

\begin{document}

\twocolumn[\hsize\textwidth\columnwidth\hsize\csname
@twocolumnfalse\endcsname

\title{Semiclassical instability of the brane-world: Randall-Sundrum bubbles}
\author{Daisuke Ida$^{1}$, Tetsuya Shiromizu$^{1}$ and 
Hirotaka Ochiai$^{2}$}
 \address{$^1$ Research Centre for the Early Universe (RESCEU),The University of Tokyo,
Tokyo 113-0033, Japan}
 \address {$^2$ Department of Physics,
The University of Tokyo,
Tokyo 113-0033, Japan}

\date{\today}

\maketitle

\begin{abstract}

We discuss the semiclassical instability of the Randall-Sundrum 
brane-world model against a creation of a kind of Kaluza-Klein bubble. 
An example describing such a bubble space-time is constructed from 
the five-dimensional AdS-Schwarzschild metric. The induced geometry 
of the brane looks like the Einstein-Rosen bridge, which connects 
the positive and the negative tension branes. The bubble rapidly 
expands and there also form a trapped region around it.

\end{abstract}
\vskip2pc]


\vskip1cm


In recent progress in string/M-theory, the brane-world scenario 
has been received much attention.  This scenario gives us a new 
possible picture of our universe. The simplest model has been 
proposed by Randall and Sundrum(RS models)\cite{RSI,RSII}. Therein 
the brane consists of four-dimensional Minkowski space-time  
located at the boundary of the bulk five-dimensional anti-de Sitter (AdS) 
space-time. It can be checked that four-dimensional gravity is 
recovered at low energy scales on the brane \cite{Tama,Tess}. 
In addition, there are  exact solutions describing the homogeneous 
and isotropic expanding universe \cite{Cosmos,Creation}. Unfortunately, 
we do not know the fundamental features of black holes so 
much \cite{BH1,BH2}. 

Although RS models has great success, there seems to be a crucial 
problem of the stability. It is well known in the standard 
Kaluza-Klein theory that the Kaluza-Klein vacuum is unstable 
against the decay channel to the so called Kaluza-Klein bubble space-time
\cite{KKB,Gary}. Accordingly, we worry about the similar instability 
in RS models. This has been firstly pointed out in Ref. 
\cite{Shinkai,Horava}(See Ref. \cite{Mark} for another instability 
in lower dimensions.). 
However, the exact solution describing the Kaluza-Klein bubble 
space-time has not been presented in the RS brane-world context.  

In this paper we will present an explicit example describing a sort of 
the Kaluza-Klein bubble(RS bubble) in two branes system in the RS brane-world 
context(RSI models). Then we show that the geometry on the brane 
has the structure of the Einstein-Rosen bridge \cite{ER}, 
which connects the positive 
and negative tension branes. Thus the solution presented here 
expresses a kind of black hole in the brane-world, though it might 
not be what we want in the low energy scales.

\begin{figure}[t]
\vspace*{-5mm}
\begin{center}
\epsfxsize=3.0in
\epsffile{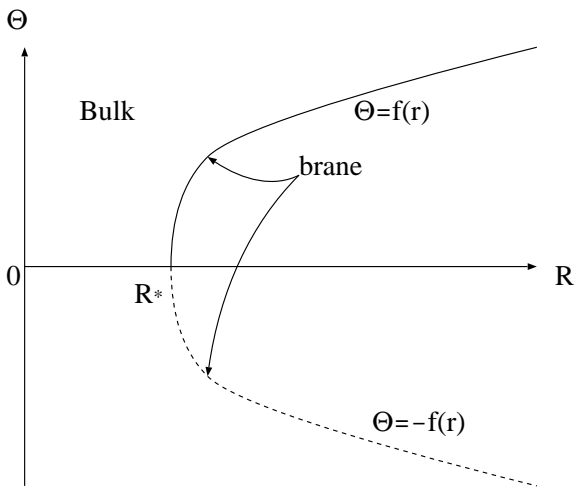}
\end{center}
\caption{The location of a brane 
in $(R,\Theta)$-plane for RSII single-brane system.}
\label{fig1}
\end{figure}

The Randall-Sundrum model of single brane system (RSII)\cite{RSII} 
is given by the metric of the form
\begin{equation}
g=dy^2+e^{-2|y|/\ell}q_{\mu\nu}dx^\mu dx^\nu,\label{RS}
\end{equation}
where $q$ is the four-dimensional Minkowski metric.
The metric (\ref{RS}) is that of of the five-dimensional AdS space,
and the brane is located at $y=0$ on which
\begin{equation}
K_{\mu\nu}:=\frac{1}{2}  \mbox \pounds_n h_{\mu\nu}=-\frac{1}{\ell} h_{\mu\nu},
\label{junction}
\end{equation}
is satisfied, where $n=\partial_y$ and $h_{\mu\nu}=e^{-2|y|/\ell}q_{\mu\nu}$
is the unit normal vector and the induced metric of a $y={\rm constant}$ 
hypersurface. If the four-dimensional metric $q$ is replaced by a Ricci-flat
metric, then Eq.~(\ref{RS}) represents an more generic Einstein metric.
Let us write the brane-metric in the form
\begin{equation}
q=-r^2d\tau^2+dr^2+r^2\cosh^2\tau d\Omega_2{}^2,\label{Rindler}
\end{equation}
where $d\Omega_2{}^2$ denotes the standard metric of the unit two-sphere.
The metric (\ref{Rindler}) represents the Rindler space, which is locally flat,
but geodesically incomplete at the null hypersurface $r=0$ (Rindler horizon).
Each $r={\rm constant}$ hypersurface corresponds to the world sphere
in a uniformly accelerated expansion.
We here consider another generalization of Eq.~(\ref{RS}) with a same 
asymptotics as Eq.~(\ref{Rindler}) on the brane.
This is given by
\begin{eqnarray}
g&=&\left[\frac{1-(\rho_*/ar)^2}{1+(\rho_*/ar)^2}\right]^2dy^2
+a^2\left[1+\left(\frac{\rho_*}{ar}\right)^2\right]^2(-r^2d\tau^2\nonumber\\
&&+dr^2+r^2\cosh^2\tau d\Omega_2{}^2),
\label{exp}
\end{eqnarray}
where
$\rho_*>0$ is a constant and $a:=e^{-|y|/\ell}$.
The metric (\ref{exp}) solves the five-dimensional Einstein equation 
with a negative cosmological term and Eq.~(\ref{junction}) is also 
satisfied at every $y={\rm constant}$ hypersurface. The coordinate 
system used here is inappropriate at $ar=\rho_*$, however this is 
only a coordinate singularity as shown below. The metric (\ref{exp}) 
is obtained by analytic continuation of the five-dimensional
AdS-Schwarzschild space-time, of which metric has the form
\begin{eqnarray}
g&=&-F(R)dT^2+F(R)^{-1}dR^2\nonumber\\
&&+R^2 (d\chi^2+\sin^2\chi d\Omega_2{}^2),\label{schads}\\
F(R)&=&1-\left(\frac{R_*}{R}\right)^2+\left(\frac{R}{\ell}\right)^2.
\end{eqnarray}
This metric can be analytically continued at the totally geodesic surfaces
$T=0$ and $\chi=\pi/2$ by replacement of the coordinates
\begin{equation}
T\mapsto i\Theta,~~~\chi\mapsto\frac{\pi}{2}+i\tau.
\end{equation}
Then the new metric becomes
\begin{equation}
g=F(R)d\Theta^2+F(R)^{-1}dR^2+R^2 (-d\tau^2+\cosh^2\tau d\Omega_2{}^2),
\label{newschads}\\
\end{equation}
which represents the straightforward generalization of the Kaluza-Klein 
bubble. The $(\Theta,R)$-plane is geodesically incomplete at
\begin{equation}
R=R_h:=\ell \left[\frac{1}{2}\left(1+\frac{4R_*^2}{\ell^2}\right)^{1/2}-\frac{1}{2}\right]^{1/2},
\end{equation}
which can be removed by making $\Theta$ periodic with the period given by
the inverse Hawking temperature: $\beta_H:=4\pi/F'(R_h)$.
We shall however temporally regard the coordinate $\Theta$ as non-periodic.
To arrive at the brane-world metric (\ref{exp}), we consider the
coordinate transformation given by
\begin{eqnarray}
R&=&ar\left[1+\left(\frac{\rho_*}{ar}\right)^2\right],\\
\Theta&=&y+\frac{1}{\ell}\int^R_{R_*}\frac{R}{F(R)}
\left(1-\frac{R_*{}^2}{R^2}\right)^{-1/2}dR,
\end{eqnarray}
where $\rho_*=R_*/2$ and $a=e^{-y/\ell}$, and the coordinates range over
$(-\infty<y<+\infty,ar>\rho_*)$.
This chart covers the region $R>R_*$ of the $\{\Theta,R\}$-coordinate 
system. If we impose the $Z_2$-boundary condition at $y=0$ surface, 
then we will obtain the brane-world model. This however is not sufficient, 
since $y=0$ surface is geodesically incomplete at 
$r=\rho_*$ [$(\Theta,R)=(0,R_*)$]. This can easily be made geodesically 
complete by reflecting with respect to the surface $\Theta=0$; 
If the $y=0$ surface is given by $B_+$: $\{\Theta=f(R)\}$, then the 
reflected surface $B_-$: $\{\Theta=-f(R)\}$ smoothly continues to $B_+$ 
at $R=R_*$. We obtain the brane-world model with the brane at 
$B=B_+\cup B_-$ in this way (see Fig.~\ref{fig1}).

However, the bulk is geodesically incomplete since it contains the point 
$R=R_h$, if a {\it single positive tension} brane is considered.
Therefore, the coordinate $\Theta$ should be periodic in this case.
Then, the brane intersects itself at a point given by 
$f(R)=\beta_H/2$, 
where a domain wall (in a four-dimensional sense) should be located.
This means that the brane has a spatially compact topology, so that this 
is not asymptotic to the RSII model. See Ref. \cite{Ruth} for the similar 
argument in the different context. 

\begin{figure}[t]
\vspace*{-5mm}
\begin{center}
\epsfxsize=3.0in
\epsffile{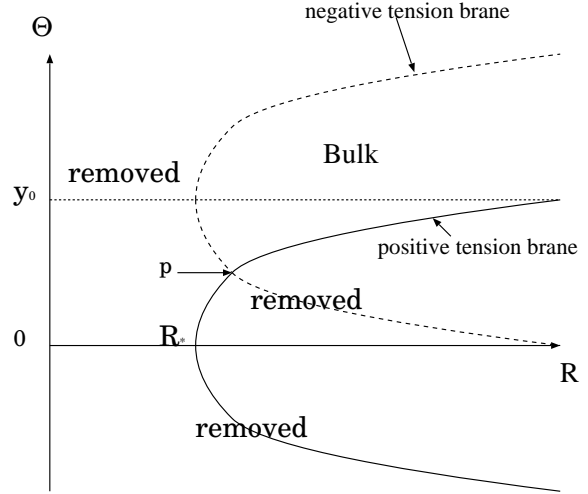}
\end{center}
\caption{The location of branes in $(R,\Theta)$-plane for RSI two-brane system.}
\label{fig2}
\end{figure}

Next, let us consider a generalization of the Randall-Sundrum model 
with two branes (RSI), in which a pair of branes with respective 
positive and negative tension is parallelly located at the boundary of 
the AdS bulk. In the present case, since the $(\Theta,R)$-plane is 
invariant under the translation in $\Theta$-direction, we can consider 
many copies of the brane already constructed by such a parallel translation. 
If the positive tension brane is given by $B=B_+\cup B_-$, the negative 
tension brane can be obtained by $\bar B=\bar B_+\cup \bar B_-$, where 
$\bar B_\pm$: $\{\Theta=\pm f(R)+y_0\}$, and $y_0$ denotes the separation of branes.
Two branes $B$ and $\bar B$ intersect at $p$ given by $\Theta=y_0/2$; namely,
two branes are connected (see Fig.~\ref{fig2}).
In the present case, we need not make $\Theta$ periodic, since the center 
$R=R_h$ can be sealed off behind the negative tension brane, so that 
we can obtain a brane-world model asymptotic to RSI.
Note that the induced metric of the brane is smooth at $p$, where
just the embedding of the boundary is singular; In fact, the intrinsic
geometry of the brane constructed here is same as that of $B$ in isolation.

\begin{figure}[t]
\vspace*{-5mm}
\begin{center}
\epsfxsize=3.0in
\epsffile{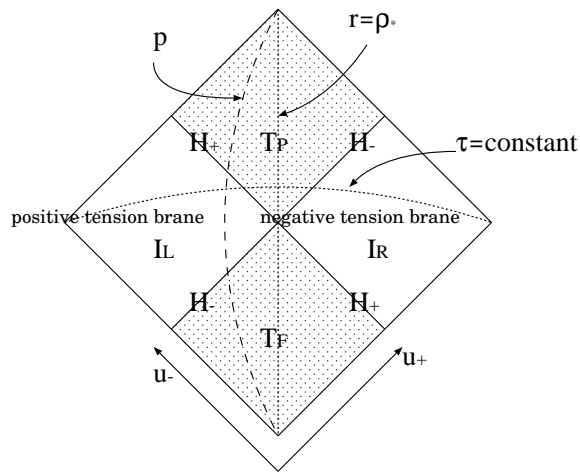}
\end{center}
\caption{The Penrose diagram for the induced geometry of the brane 
corresponding to the situation in FIG. 2. 
The dashed line denotes the surface $p$ connecting two branes.}
\label{fig3}
\end{figure}

Here we shall consider the induced metric $h$ of the brane $B$.
It can be shown that $h$ is given by
\begin{equation}
h=\left[1+\left(\frac{\rho_*}{r}\right)^2\right]^2(-r^2d\tau^2+dr^2+r^2\cosh^2\tau
d\Omega_2{}^2).
\label{branemetric}
\end{equation}
The coordinate $r$ now ranges all positive value, where the region $r>\rho_*$ 
corresponds to $B_+$ and $0<r<\rho_*$ to $B_-$ [note that the metric (\ref{branemetric}) 
is invariant under $r\mapsto\rho_*^2/r$]. Let us introduce null coordinates 
$u_\pm=\tau\pm\ln (r/\rho_*)$, then the metric (\ref{branemetric}) becomes
\begin{eqnarray}
h&=&-\rho_*{}^2e^{u_++u_-}\left(e^{-u_+}+e^{-u_-}\right)^2du_+du_-\nonumber\\
&&+{\cal R}(u_+,u_-)^2d\Omega_2{}^2,
\end{eqnarray}
where
\begin{equation}
{\cal R}(u_+,u_-)=\frac{\rho_*}{2}(1+e^{u_++u_-})(e^{-u_+}+e^{-u_-}).
\end{equation}
Then, the expansion rates of the outgoing and the ingoing spherical rays are 
given by
\begin{equation}
\theta_\pm:=\frac{\partial\ln {\cal R}}{\partial u_\pm}
=\frac{e^{u_\pm}-e^{-u_\pm}}{(1+e^{u_++u_-})(e^{-u_+}+e^{-u_-})},
\end{equation}
respectively.
There are null hypersurfaces $H_\pm$
\begin{equation}
H_\pm: u_\pm=0
\end{equation}
 on which $\theta_\pm$ vanishes, respectively.
The brane $(B,h)$ is divided by $H_\pm$ into
four regions;
(i) $I_R$: right asymptotic region
$u_+>0$, $u_-<0$ [$(\theta_+,\theta_-)=(+,-)$],
(ii) $I_L$: left asymptotic region
$u_+<0$, $u_->0$ [$(\theta_+,\theta_-)=(-,+)$],
(iii) $T_P$: past trapped region
$u_+>0$, $u_->0$ [$(\theta_+,\theta_-)=(+,+)$],
(iv) $T_F$: future trapped region
$u_+<0$, $u_-<0$ [$(\theta_+,\theta_-)=(-,-)$].
The Penrose diagram is depicted in Fig.~\ref{fig3}.

Each $\tau={\rm constant}$ hypersurface has the Einstein-Rosen 
bridge around $r=\rho_*$. Thus, both of the bulk and the brane has 
non-trivial topology
(simply connected but with non-vanishing second Betti number),
which represents the creation of a sort of bubbles({\it RS bubble}).
One however cannot traverse from one side to the other;
Once someone  steps into $T_P$, he
will never make an exit. Thus, the region $T_P$ is a kind of black holes, though there
is much difference from what we know of black holes. In particular, the total
gravitational energy vanishes, which indicates that the RSI model
might decay by creating RS bubbles semiclassically. 
A creation of a bubble implies a cross-linking of two branes through
a topology changing process of the bulk and the brane. 
There is negative energy distribution around the RS bubble, which comes from 
the electric part of the five-dimensional Weyl tensor 
\begin{equation}
E_{\mu\nu}={}^{(5)}C_{\mu\alpha\nu\beta}n^\alpha n^\beta
\end{equation}
through the effective Einstein equation on the brane\cite{Tess}
\begin{equation}
{}^{(4)}G_\mu^\nu=-E_\mu^\nu=\frac{4\rho_*^2r^4}{(r^2+\rho_*^2)^4}
(\delta^\tau_\mu\delta^\nu_\tau
-3\delta^r_\mu\delta^\nu_r
+\delta_\mu^\vartheta\delta^\nu_\vartheta
+\delta_\mu^\varphi\delta^\nu_\varphi).
\end{equation}
The energy density observed by $r={\rm constant}$ observer therefore becomes
\begin{equation}
\epsilon=-\frac{\rho_*^2r^4}{2\pi G_4(r^2+\rho_*^2)^4}<0,
\end{equation}
of which amplitude peaks at $r=\rho_*$ with $|\epsilon|=(32\pi G_4\rho_*^2)^{-1}$,
and rapidly dumps as $1/r^4$ $(r\rightarrow+\infty)$.

Finally, we shall estimate the semiclassical decay probability of the 
RSI brane-world 
using the euclidean path integral. The corresponding euclidean bounce 
solution is obtained by the Wick rotation, $\tau \to i \tau_E +\pi/2$, 
of the metric of Eq (\ref{exp}). The decay occurs at $\tau=0$ because the 
4-dimensional surfaces at $\tau=0$ is momentary static. As a result the 
decay probability will be order 
$P \sim \exp (- \frac{\ell \rho_*^2}{G_5})$.
In the above $G_5$ is the five-dimensional gravitational constant having the 
relation with the four-dimensional gravitational constant, $G_4$, as 
$G_5 \sim  \ell G_4 e^{2y_0/\ell}$. $y_0$ is the typical 
coordinate distance between two branes. In the RSI models we often 
assume $G_5 \sim 1{\rm TeV}^{-3}$ and 
$\ell \sim 1{\rm mm}$. For $\rho_* >(G_5/\ell)^{1/2}\sim 
(10^{11}{\rm GeV})^{-1}$, this decay process might be suppressed.  


Let us summarise our study. We presented an explicit example
describing the brane-world after the Randall-Sundrum models decays. 
We called this the Randall-Sundrum bubble spacetimes.    
The brane geometry has the structure of the Einstein-Rosen bridge, but 
not black hole due to the negative effective energy from the bulk Weyl 
tensor. It turns out that RSI-type models is realised in the present 
procedure, but RSII type models is not. The decay probability of RSI 
models to RS bubble spacetimes 
was roughly evaluated and we saw that the decay process crucially 
affects the RSI brane-world scenario. Supersymmetry may be important 
so that it might forbid this decay process 
in the brane-world context as well as in the standard Kaluza-Klein 
theory \cite{KKB}.

\section*{Acknowledgments}

We thank Y. Shimizu for useful discussions. HO would like to thank K. Sato 
for his continuous encouragement. TS's work is partially supported by Yamada 
Science Foundation.

\end{document}